# Skin Controlled Electronic and Neuromorphic Tattoos

Dmitry Kireev[1,2*], Nandu Koripally[1,3], Samuel Liu[1], Gabriella Coloyan Fleming[4], Philip Varkey[1], Joseph Belle[1], Sivasakthya Mohan[5], Sang Sub Han[7], Dong Xu[8], Yeonwoong Jung[7,9,10], Xiangfeng Duan[11,12], Jean Anne C. Incorvia[1], Deji Akinwande[1,6,13]

[1] Chandra Department of Electrical and Computer Engineering, The University of Texas at Austin, Austin, TX, USA

[2] Department of Biomedical Engineering, University of Massachusetts Amherst, Amherst, Massachusetts, MA, USA

[3] Department of Electrical and Computer Engineering, University of California San Diego, La Jolla, CA, USA.

[4] Walker Department of Mechanical Engineering, The University of Texas at Austin, Austin, TX, USA

[5] Department of Materials Science and Engineering, The University of Texas at Austin, Austin, TX, USA

[6] Texas Materials Institute, The University of Texas at Austin, Austin, TX, USA

[7] NanoScience Technology Center, University of Central Florida, Orlando, FL, USA

[8] Department of Materials Science and Engineering, University of California Los Angeles, Los Angeles, CA, USA

[9] Department of Materials Science and Engineering, University of Central Florida, Orlando, FL, USA

[10] Department of Materials Science and Engineering, University of Central Florida, Orlando, FL, USA

[11] Department of Chemistry and Biochemistry, University of California Los Angeles, Los Angeles, CA, USA

[12] California NanoSystems Institute, University of California Los Angeles, Los Angeles, CA, USA

[13] Department of Biomedical Engineering, The University of Texas at Austin, Austin, TX, USA

*Corresponding author. Email: dkireev@umass.edu

**Abstract:** Wearable human activity sensors developed in the past decade show a distinct trend of becoming thinner and more imperceptible while retaining their electrical qualities, with graphene e-tattoos, as the ultimate example. A persistent challenge in modern wearables, however, is signal degradation due to the distance between the sensor's recording site and the signal transmission medium. To address this, we propose here to directly utilize human skin as a signal transmission medium as well as using low-cost gel electrodes for rapid probing of 2D transistor-based wearables. We demonstrate that the hypodermis layer of the skin can effectively serve as an electrolyte, enabling electrical potential application to semiconducting films made from graphene and other 2D materials placed on top of the skin. Graphene transistor tattoos, when biased through the body, exhibit high charge carrier mobility (up to 6500 $cm^2V^{-1}s^{-1}$), with $MoS_2$ and $PtSe_2$ transistors showing mobilities up to 30 $cm^2V^{-1}s^{-1}$ and 1 $cm^2V^{-1}s^{-1}$, respectively. Finally, by introducing a layer of Nafion to the device structure, we observed neuromorphic functionality, transforming these e-tattoos into neuromorphic bioelectronic devices controlled through the skin itself. The neuromorphic bioelectronic tattoos have the potential for developing self-aware and stand-alone smart wearables, crucial for understanding and improving overall human performance.

It has been shown that 2D materials are unique elements that bear significant potential for developing next-generation wearable bioelectronics[1–4]. However, there are two limiting factors that slow down the penetration of 2D materials into scalable usage in wearables. The first factor is the quality of 2D materials and the dimensions of their crystals. Even the highest quality chemical vapor deposition (CVD) grown graphene, commercially available, has grain sizes of ~100 µm. The grain boundaries in graphene, however, are not highly crucial due to the metallic/semi-metallic nature of the graphene. Hence, large-scale graphene devices have been made up to a few cm long. On the other hand, the semiconducting 2D materials, such as $MoS_2$, $WS_2$, and $WSe_2$, suffer significantly from the presence of grain boundaries, and creating cm, even mm-scale devices is still challenging[5]. The second limiting factor is the lack of studies of the physical and electronic properties of 2D materials in an ambient environment. This is partially due to the environmental instability of other 2D materials like black phosphorus and silicene[6,7].

Recently, Zhao *et al.* used the skin as a medium for ionic communication[8]. They proposed an intra-body data transfer approach that uses ion-conducting tissue as a data-transporting media. While operating at a high-frequency range, they note that at frequencies below 1 kHz, the electrical double layer regime will be dominant, which is well suited for many applications such as the one shown in this work. A somewhat similar strategy was recently reported by Yan *et al.*[9] where skin and plant tissues were used to apply field effect onto the free-standing membrane of sub-mm size semiconductor flakes. In contrast, here, we show that the skin can host and effectively activate cm-scale semiconductor channels such as graphene.

In this work, we report our experimental finding on a unique system that can be used to study 2D material's properties in a mechanically soft and flexible environment that is more realistic for future wearable applications. The reported system consists of commercially available medical-grade gel electrodes, which contain everything that one requires to apply an external electric field to a semiconductor: solid gel electrolyte and Ag/AgCl reference electrode. We thoroughly tested five different gel electrodes and six different graphene transistor tattoos, drawing essential conclusions for wearable applications. Then, we translated the technology into the human body. We hypothesized that electrochemical gating of the **graphene field effect transistors tattoo (GFET-T)** is possible by applying field-effect potential right through the skin (**Fig. 1**). The experiments were safely performed on porcine skin and human skin consecutively, compliant with all safety conditions. Noteworthy is the field-effect mobility of our body-gated graphene transistors, which reached up to ~6500 $cm^2V^{-1}s^{-1}$, $MoS_2$ transistors up to 30 $cm^2V^{-1}s^{-1}$ and $PtSe_2$ up to 1 $cm^2V^{-1}s^{-1}$. Finally, we took the technology one step further and hypothesized that the skin could be used for neuromorphic functions, similar to the EC-RAM devices[10]. By adding a layer of Nafion, we show that such neuromorphic functions can also be expanded for body-body communication, showing remarkable (near-ideal) inter-body neuromorphic computing.

## Main

**Gel as a low-cost and reliable solid electrolyte interface for FET characterization**

The reported method relies on the technology of ultrathin graphene-based electronic tattoos (GETs)[11–13]. Developed earlier, the graphene tattoos have been used as passive interfaces exhibiting unmatched mechanical and electrical interfaces with the human body[14]. The 1 atomic layer (1L), 2L, and 3L graphene tattoos with and without holes (micro-embossed into the structure of graphene tattoos, see Fig. S1) were fabricated and transferred on the skin and on skin-like hydrogels with a set of the source and drain electrodes (**Fig. 1a**). Fabrication of the six kinds of graphene tattoos followed the latest "GETs 2.0" protocol[15,16]. The commercial Ag/AgCl gel electrodes are used here as a reference for on-skin testing (Fig. S2), but also as a stand-alone test-bed, making a plausible case for the application, since the embedded Ag/AgCl material is routinely used as a proper electrochemical reference in FETs (Fig. S3). In the case of skin-gated devices, the additional reference electrode (same Ag/AgCl gel) is applied laterally *via* the skin (**Fig. 1**). Just as hypothesized, the devices work well *via* gel, where charged ions within the electrolyte, conductive gel, or skins' dermal layer are responsive to the applied potential (**Fig. 1b**), forming an electrical double layer (EDL). The measurements are performed in DC mode, and the sweep rate is slow (40 mV/sec). In this mode, the electrophysiological signals are not directly affecting the device performance. As one can see from an aggregate **Fig. 1c**, the GFET-Ts operate when gated *via* gel, porcine skin, and human skin.

Before testing the devices on human skin, we extensively characterized the performance of the gel devices in order to ensure safe operation parameters. To conduct a systematic study, we evaluated five different gel electrodes, namely 3M models 2330, 2360, 2560, and 2670-5 from 3M and Meditrace from Cardinal Health (Fig. S4).

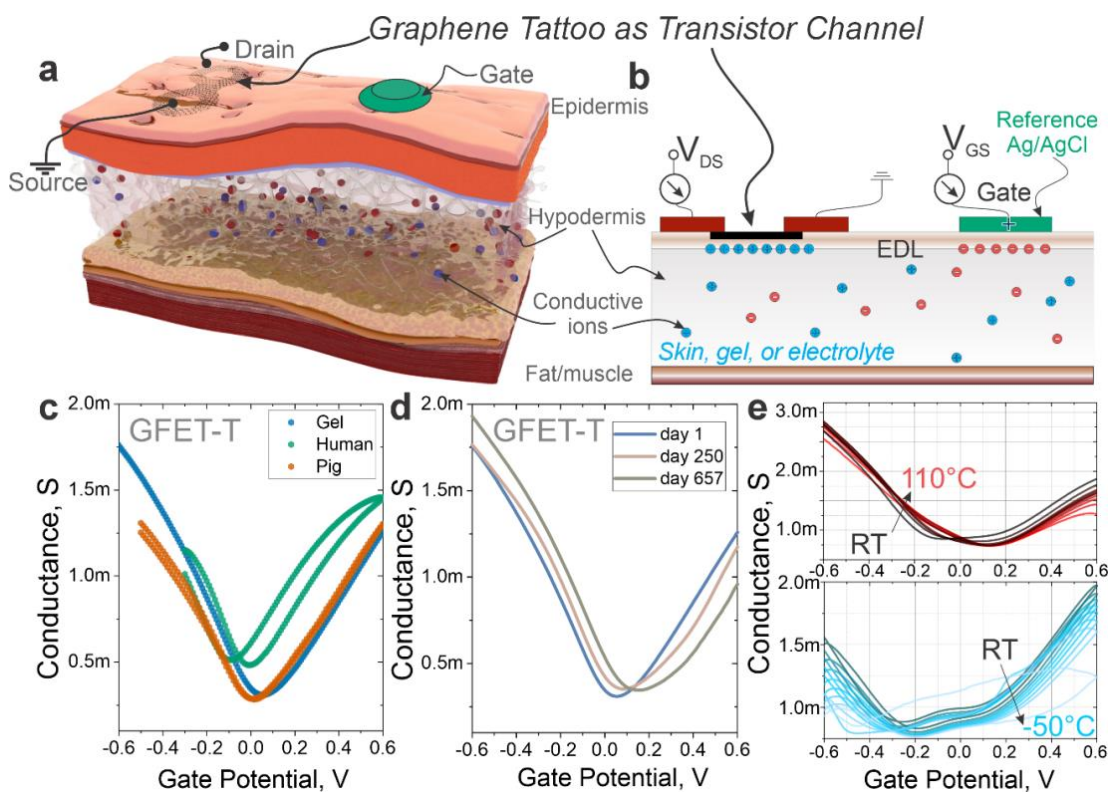

**Fig. 1 | Skin as a functional bioelectronics interface**. **a**, Schematics of the skin with its structural elements (epidermis, hypodermis with conductive free ions, fat, and muscle), with hypodermis active as electrolyte. **b**, A unified schematic equalizing gel-gating as well as skin-gating of a graphene tattoo transistor, with hypodermis (or solid electrolyte) active as a media with fee ions, and gate potential is applied on a side *via* an Ag/AgCl electrode. **c**, GFET-T performance curves measured through the gel (blue), porcine skin (orange), and human skin (green). **d**, Transfer curves of a 2L-GFET-T measured *via* gel #2330 for >2 years at ambient. **e**, Transfer curves of two GFET-Ts measured *via* gel #2360 while heating up (top, shaded from black (RT) to red (110°C)) and cooling down (bottom, shaded from black (RT) to light blue (-50C°)).

We first examined their electrochemical properties, focusing on the interface capacitance, which is a critical factor for device performance. As shown in Fig. S4, there is a significant variation in capacitance across the different gels, likely influenced by the differences in their composition, thickness, and the Ag/AgCl reference material used in each electrode. For instance, the thickness of the gels varied significantly, with the #2330 and #2360 models measuring ~300 μm, #2560 at 450 μm, #2670-5 at 235 μm, and Meditrace at 1200 μm. These thickness variations, combined with the different backing materials and adhesive properties, likely contributed to the observed discrepancies in capacitance values. In terms of performance, the #2330 and #2360 models yielded the most reproducible capacitance values, and are constructed with a "hydrogel polymer", which could also explain their consistent performance. In contrast, the model #2560 uses polyethylene glycol dimethacrylate as its conductive adhesive.

Considering the electrochemical performance, we assume that the interface capacitance of the #2330 and #2360 models is closer to the electric double layer capacitance typically observed in highly concentrated electrolytes, approximately 1.4 µF/cm² as referenced in previous studies[17–19]. Finally, at the voltages near the Dirac point, we anticipate the additional effect from the graphene's quantum capacitance; however, in this work, we report on the single highest values of mobilities that are usually far from the Dirac point[19].

We have found that the gel electrolytes are robust towards mechanical strain, feature small leakage currents, and are highly stable over time. Contrary to the common belief that the gel electrodes tend to dry and fail within weeks, this is not the case for all. Especially, the #2330 and #2360 from 3M are ultra-robust; as one can see from **Fig. 1d**, the same device's performance did not degrade even after >2 years of storage at ambient conditions. The other gels dried out after ~3 months in ambient. Finally, as shown in **Fig. 1e**, heating from room temperature (RT) up to 110°C or cooling down to -50°C does not significantly affect the GFET-T's properties, suggesting a rather broad operating temperature range. The gels are also stable in Ar atmosphere (Fig. S5), as well as thy do not degas and do not contaminate the oxygen-free gloveboxes – an essential advantage of working with other 2D materials that must be stored in an air-free environment.

Then, we focused on characterizing the performance of 6 different GFET-T compositions (1L, 2L, and 3L graphene, each with and without holes) on 5 different gel types, with 5 devices per each configuration (N=150 samples in total). Micro-sized (~100 µm) holes were embossed in the graphene tattoos to enhance their water vapor transmission (Fig. S1). As previously reported[16], the holed graphene tattoos feature an impressive water vapor transmission rate (WVTR) of 2770±494 g/m²/day. The overall performance of the GFET-Ts of different compositions and measured with different gel electrodes have little similarities. Even the simplest figure-of-merit – location of the charge neutrality point (CNP or Dirac point) highly depends on the gel type used (Fig. S6-7), indicating that the gel itself applies additional background doping to the graphene. However, even when using the same gel (*e.g.,* 2360, Fig. S7), the CNP shifts depending on the GFET-T structure and number of layers. The transconductance trend (Fig. S8) shows the most expected performance, which is saturated at the 2L and 3L graphene samples regardless of the gel electrode. The large discrepancy in the 1L devices' performance, especially in the 1L holey (1Lh) ones, is due to the structural inhomogeneity of the monolayer graphene. Finally, the overall on/off ratio slightly decreased with the increasing number of graphene layers in the stack (Fig. S9), which is also expected, although the statistical analysis shows that the groups are not significantly different (p>0.05). The primary conclusion from these tests is that using 2L-GFET-Ts is preferential since they provide superior performance to monolayers; while no substantial improvement has been reported in 3L devices.

**Rapid probing of 2D transistors**

Measuring the performance of a pristine 2D material without transfer is a critical gap that has not been entirely addressed up to now. Here, we show that by using gel electrodes, we can achieve an unprecedented level of device characterization in terms of yield and simplicity. The proposed gel devices can be used as a rapid and low-cost characterization test bed for prototyping and rapid probing of 2D materials and their field-effect performance.

As a proof of concept, besides characterizing CVD-grown graphene, transferred from copper foils and supported by poly(methyl methacrylate) (PMMA), we tested three other kinds of 2D materials: $MoS_2$, $PtSe_2$, and graphene grown on sapphire (see Methods for details). The materials were chosen to highlight the robustness of our testbed system: it can measure all sorts of 2D materials: either directly grown on a rigid substrate, grown on a flexible substrate, or spin-coated on an arbitrary substrate. Since the $MoS_2$ and $PtSe_2$ have a rather short effective channel length, the 5 mm channel length used for GFET-Ts is not suitable; hence we evaporated the source and drain conductive channels *via* a shadow mask, creating a device with an effective channel length of 50 μm and W/L ratio of ~500 but the footprint of ~5 mm² (Fig. S10). **Fig. 2a** shows the schematic of the experimental routine, highlighting the ability to measure the performance of the as-grown 2D materials (on arbitrary and non-dielectric substrates), since transfer of 2D materials, especially at large scale is the major cause of performance degradation, cracks, and defects[20,21]. The I-V (transfer) curves of the $MoS_2$ and $PtSe_2$ transistors (**Fig. 2b,c**) show that their performance (mobility) is on par with the expectations based on the state-of-the-art. $MoS_2$ features mobility ~30 cm²V⁻¹s⁻¹. The $MoS_2$ transfer curve shows saturation at +1 V, and negligible drain-source current at 0 V and below. The gate leakage current is negligible, as it is at least 2-3 orders of magnitude below the drain-source currents (Fig. S11 for details). Purity of the $MoS_2$ flakes has been thoroughly confirmed prior to use (see Methods)[22]. As a way towards highly integrated wearable amplification circuits, we went beyond $MoS_2$ to another 2D material that can also be grown at rather low temperatures: $PtSe_2$. The $PtSe_2$ sample is showing ~1 $cm^2V^{-1}s^{-1}$ mobility, which is also expected since the $PtSe_2$ was grown on a large scale, in the thermal assisted conversion (TAC) process[23], and it is actually one of the largest values measured for this kind of material (see Methods). Finally, we also measured another kind of graphene – grown directly on sapphire, which required careful transfer-free characterization of the properties **(Fig. 2d)** since the transfer from sapphire is typically extremely challenging. The graphene-on-sapphire pieces were simply diced into ~5x10 mm pieces and placed on top of the gel test-beds graphene down for characterization. Very important to note here that measured mobility of graphene-on-sapphire samples ~680 $cm^2V^{-1}s^{-1}$ is below the direct CVD-grown graphene. This is due to the quality of the growth that results in high-density small crystal size bilayer graphene[24].

Overall, this is a convenient no-transfer stick-and-measure approach for one-step integration and characterization of the Van der Waals materials with atomically thin interfaces.

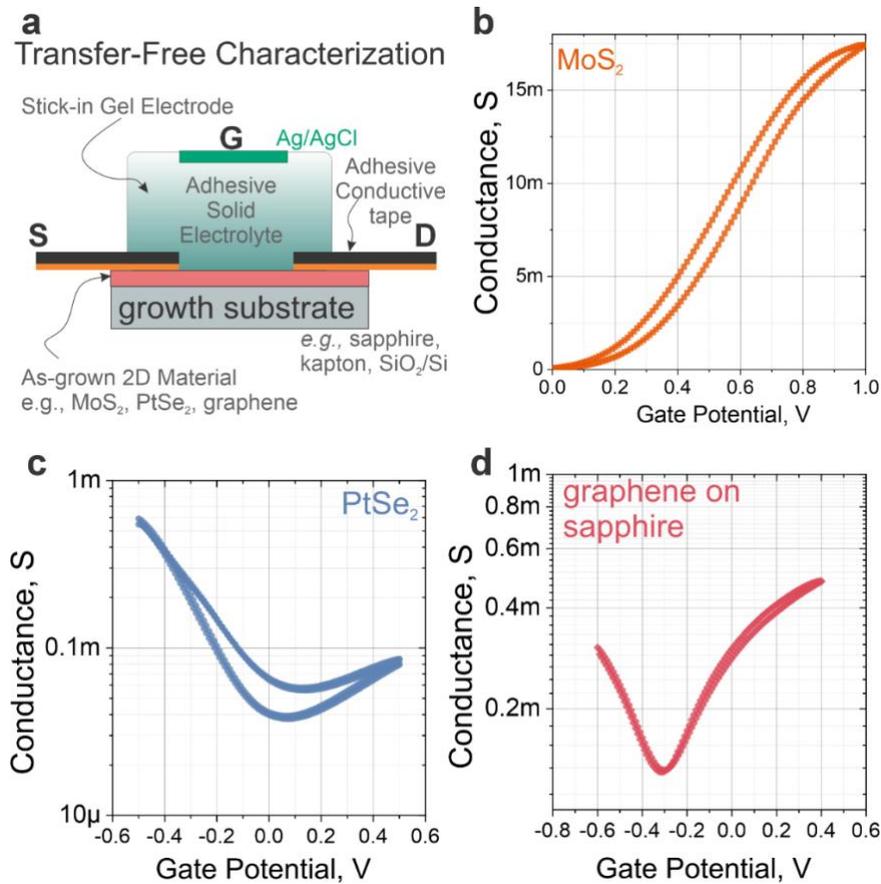

**Fig. 2 | Transfer-free characterization of 2D materials**. **a**, schematic of the set-up that employs any 2D material (red) grown on an arbitrary insulating substrate (gray), contacted by

adhesive conductive tape to form source (S) and drain (D) elements, and gate potential applied via gate reference electrode that is part of the adhesive and non-damaging. **b-d**, examples of different 2D materials *via* gel test-bed without any transfer. The transfer curves are shown for few-layered $MoS_2$ (**b**, orange) showing maximum mobility ~50 $cm^2V^{-1}s^{-1}$, few-layered $PtSe_2$ (**c**, blue) showing maximum mobility ~1 $cm^2V^{-1}s^{-1}$ and bilayer graphene-on-sapphire (**d**, red) showing maximum mobility ~680 $cm^2V^{-1}s^{-1}$.

## Skin as electrolyte.

**Porcine skin**. Upon testing of the transistors on gel electrodes, we validated our hypothesis on the porcine skin (**Fig. 3a-b** for GFET-Ts and **Fig. 3c** for $MoS_2$-FETT) followed by proof-of-concept experiments on human skin (**Fig. 3d-g**). Similar to the gel tests, we tested 6 different kinds of graphene on porcine skin experiments, with a total N=55 devices, and measured their performance. The evaluated transconductance is shown in Fig. S12. The interface capacitance was considered to be similar to the case of diluted PBS, ~1.4 µF/$cm^2$ [19]. It is important to note that the tests were performed in highly controlled settings in order to ensure that no currents above 50 µA are passed through either of the terminals to ensure safe operation during the next stage of human-gating[25].

Furthermore, we evaluated the practical aspects of the skin-gated GFET-T performance. First, the porcine skin-gated GFET-Ts were found stable for >10 days of measurements (Fig. S13). Since the Ag/AgCl gel electrodes are used as a reference, we've tested whether their kind (gel types, Fig. S14a), dimension (10–400 $mm^2$, Fig. S14b), or distance (45–130 mm, Fig. S14c) affects the device performance. Clearly, none of these qualities affect the GFET-T performance, suggesting that any further fluctuation of the properties will result from a physiological, skin, or body function variation, when connected in according (not DC) mode.

Finally, the mobility of the six graphene types measured through porcine skin was analyzed. As one can see from **Fig. 3b**, it follows the same trend as for gel devices. Clearly, the 1L devices perform the worst in terms of actual mobility but also in their standard deviation. The addition of holes almost diminishes the performance of 1L-GFET-Ts but does not significantly affect the performance of 2L-GFET-Ts and 3L-GFET-Ts. Furthermore, 3L devices do not offer any further significant improvement in performance, but they do require more effort in fabrication; hence, we elect 2L-GFET-Ts as the most suitable material for further use.

Similar to GFET-Ts, $MoS_2$-FET-Ts can be placed on porcine skin and characterized (**Fig. 3c**), showing classic $MoS_2$-specific field-effect performance, on par with the classic characterization approaches[26].

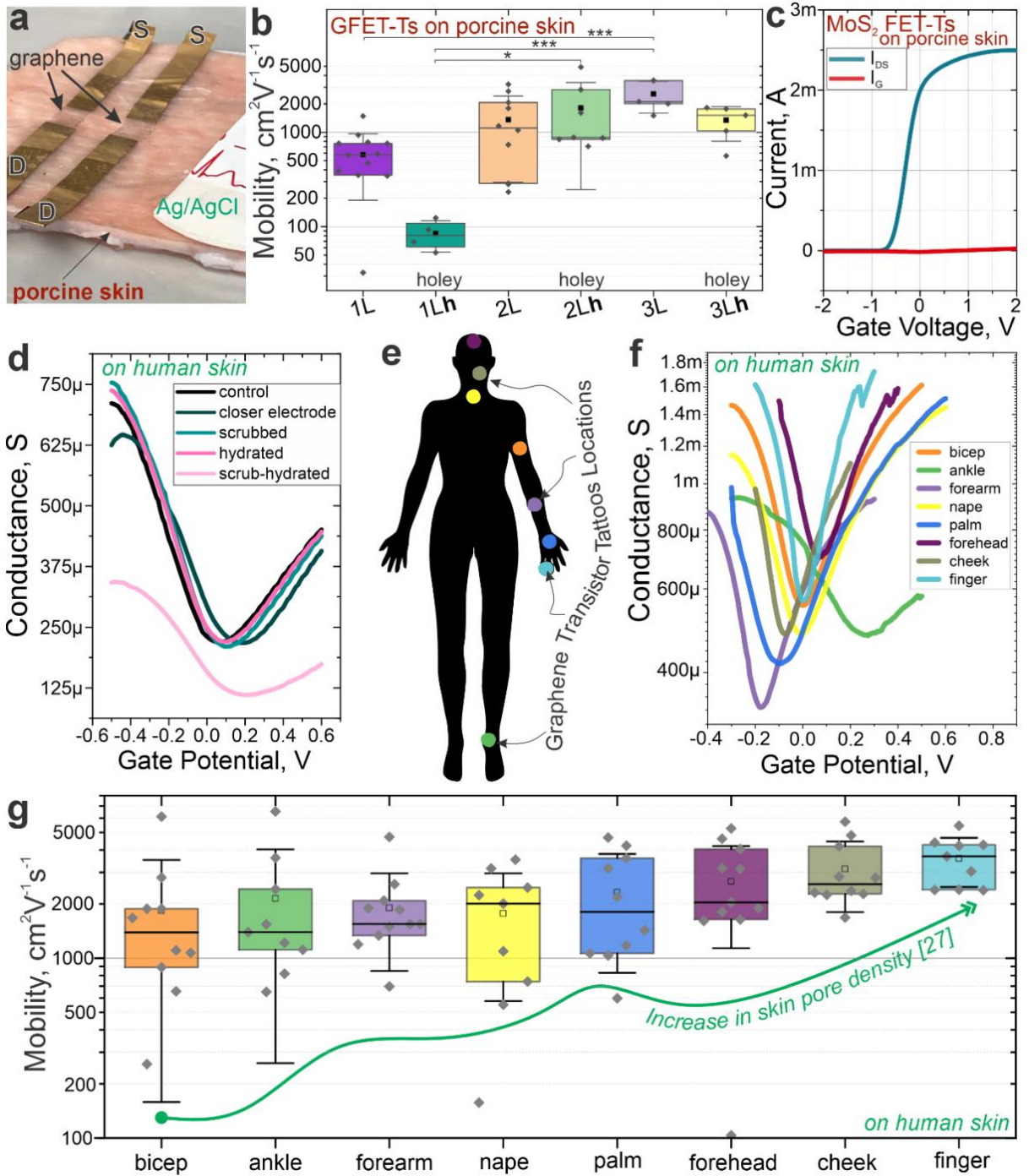

**Fig. 3 | Porcine and human skin as functional interface. a**, Photograph of two GFET-Ts on a porcine skin with two pairs of Source (S) and Drain (D) terminals made of soft conductive adhesive tape. Graphene is placed facing down the skin. Ag/AgCl gel reference electrode is used as a gate. **b**, Statistical distribution of mobility measured from 6 different kinds of GFET-Ts at least 6 of each kind. The box represents 25% and 75% with a mean; outliers are ±SD. **c**, Transfer curves of a porcine skin-gated MoS$_2$ FETT. **d**, Transfer curves of a human skin gated GFET-T measured with different conditions including skin scrubbing ("scrubbed"), lotion hydration

("hydrated"), or both ("scrub-hydrated"), and changing the Ag/AgCl electrode location. **e-f**, Schematic of the 8 different body GFET-T placement and transfer curves of the identical 2L-GFET-Ts from a single subject. **g**, Statistical distribution of the mobility measured from 8 body locations and N=12 subjects. The color code is associated with location, same as in **e-f**. The box represents 25% and 75% with a mean; outliers are ±SD. The groups are not significantly different (p>0.05). A subject-specific graph is shown in Fig. S15.

**Human skin**. Finally, for the performance validation on human skin, we only utilized here the 2L graphene structures as they were found to be optimal in terms of performance and performance variability [13]. First, alike to the study with porcine skin, we verified that the distance between reference electrode (5cm *vs* 10 cm distance) placement does not affect the measurements (**Fig. 3d**) which is easily explained by the fact that hypodermis is highly conductive liquid, only requiring a stable interface to the Ag/AgCl electrode to enable formation of electrical double layer. Likewise, we verified whether additional skin scrubbing (with Nyprep abrasive gel) or additional hydration (with Neutrogena Hydro Boost gel) would not significantly affect the device performance. The only significant change in performance was seen when the skin was consequently scrubbed and hydrated, which could be assigned to creating excessively oily surface, preventing gating efficacy (**Fig. 3d**).

Characterizing the performance of the identical 2L GFET-Ts placed on 8 different body locations for 12 subjects (some subjects did not go through all body part measurements; Table S2), we found an interesting trend in the recorded transfer curves (**Fig. 3e-f**), indicative of both intrinsic doping to graphene (shift of CNP) and different transconductance and consequently, mobility (**Fig. 3g** and Fig. S15 for individual-colored data points). While the increase in mobility values may somewhat be associated to the density of sweat glands of different skin areas[27], the hypothesis does not hold for the palm and the finger. of sweat glands, affecting the interfacial capacitance and measured values. Moreover, statistical analysis shows that the data is not significantly different (p>0.05). The poor correlation might be explained by the subject-to-subject variability as well as the small footprint (0.25 cm$^2$ total area) of the tattoo patches.

Benchmarking this performance against other works, we would like to note that this is the first time that skin-conformable and large-channel devices have been used for direct skin interfacing and body-gating. While there have been other attempts to use skin-conformable transistors, they were mainly focused on organic semiconductors such as CYTOP:DNTT[28], PEDOT:PSS[29], P3HT[30],

N2200[31], and p(g2T)[32,33], and even flake-based $MoS_2$ films[9]. As shown in Supplementary Table S1 and Fig. S16, GFET-Ts outperform the state-of-the-art by at least 2 orders of magnitude.

## Synaptic Skin and Inter-Body Communication

In order to show the proof-of-concept application where skin-controlled electronics will make a noteworthy impact, we chose to expand the functionality to neuromorphic applications. There has been a recent concept of energy-efficient biocompatible bilayer graphene-based artificial synaptic transistors (BLAST)[34]. Here, we modified the BLASTs into skinBLASTs, arranging the graphene-Nafion mixture into wearable and skin-conformal format (see Methods). **Figure 4a** provides schematics of the device with an optical picture at the zoom-in. Addition of Nafion layer creates an superfluous solid electrolyte that works as a gate insulator for the graphene transistor, and the slow dynamics of the ion movement within the Nafion membrane creates a synaptic effect that can be described as short-term or long-term memory or potentiation[35,36], like shown in **Fig. 4b-e**. To fully characterize the devices performance, a ramp test was performed by applying alternating 20 positive and 20 negative polarity current pulse trains of 10 µA amplitude and 5 ms duration (**Fig. 4b**). From the ramp test, the linearity and symmetry of the devices were characterized as well following the methodology described in Ref.[37] resulting in non-linearity parameter of +2.394 in the positive conductance change and -1.697 in the negative conductance change for the device characterized on human skin and +0.87/-1.70 for the device characterized on porcine, shown in **Fig. 4c**. Though both non-linearity and asymmetry is higher for the human-based device, this can be attributed by the larger conductance range, where the maximum conductance is 147% higher than the minimum conductance for the human-based device compared to 54% higher for the porcine-based device. There is a number of factors that affect the magnitude of conductance change per pulse, such as moisture of skin, adhesion, graphene quality, which resulted in the difference in conductance range. Overall, the nonlinearity and asymmetry are worse than for the BLASTs, but the results are still competitive with other memory types[38,39]. The change in conductance was characterized in response to the current pulse amplitude at a fixed pulse duration of 10 ms. The result shows a proportional response to an increase in pulse amplitude, where a negative current pulse results in a positive conductance change, and vice versa (**Fig. 4d**).

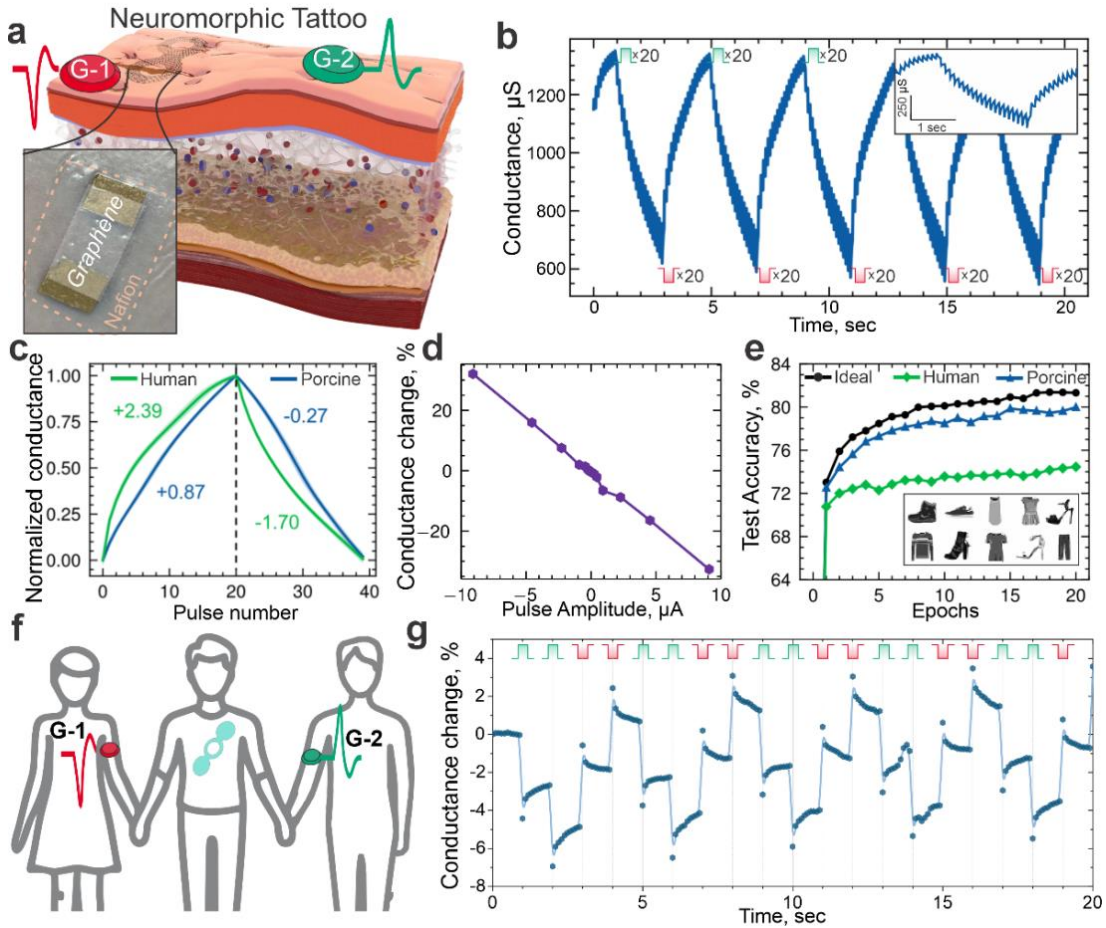

**Fig. 4 | SkinBLASTs as neuromorphic devices biased through the skin**. **a**, Schematic of one skinBLAST and two gates (marked as red and green). **b**, Performance for alternating positive and then negative trains of 20 write pulses each of 10 μA amplitude and 5 ms duration. An inset shown a 4-sec long zoom-in featuring individual steps as a response of each gate pulse. **c**, Mean conductance per pulse number measured on human skin (green) and porcine skin (blue) and associated nonlinearity values. **d**, The change in conductance (ΔG) with varied pulse amplitude (pulse duration is constant, 10 ms). **e**, Fashion-MNIST (logo shown at the inset) simulation results of the skinBLAST crossbar using the experimental data. **f**, Schematic of a multi-subject test where one subject wears the skinBLAST (middle), 1st subject has the negative gate (G-1, red), and 2nd subject has the positive gate (G-2, green). **g**, Performance of the test when the subjects hold hands showing clear inter-person 'communication' expressed as a consequence of negative (G-1, red) and positive (G-2, green) pulse trains. Additional combinations of gate pulse trains and sequences when subjects do not hold hands – see in Fig. S17.

To evaluate the performance of online learning in analog accelerators using these devices, both the porcine and human-based device data were applied to neural network simulations using the CrossSim crossbar simulator. The network used is a multilayer perceptron with 300 hidden units and applied to image recognition using the Fashion-MNIST clothing article dataset, a standard benchmark. The learning rate used for the simulation is 0.0001 with a batch size of 1. The results are shown in **Fig. 4e** and compared with those of an equivalent network composed of 64-bit ideal weights. The porcine-based network performs closer to ideal, with a more significant performance degradation seen in the human-based network. This is expected due to the higher non-linearity and asymmetry displayed in the human-based device characteristics. However, this can potentially be alleviated by reducing the operation range, resulting in more linear device characteristics. Wearable neuromorphic devices have been on the rise[40–43], while the common approach is to simply re-shape the existing electronic components into a wearable form-factor, while our approach differs by leveraging the body/skin as the biological media to transmit and eventually regulate the neuromorphic functionality. Important to note here that the number of synaptic transistors required to perform the training is $785 \times 300$ devices in the first synaptic layer and $301 \times 10$ in the second synaptic layer. While the devices in this work are not scaled, we have conducted previous work with devices on the scale of tens of micron can be fabricated with photolithography[35,36]. Other works have shown that electron beam lithography can be used to pattern Nafion on the scale of 100 nm[44]. Hence, we believe that the eventual scaling of these wearable neuromorphic networks is reasonable.

Finally, with the vision to enable inter-body neuromorphic sensing or communication (**Fig. 4f-g**), we performed a test where only one control subject wore the skinBLAST, but no gate electrodes. The two gate electrodes were actually attached to subject #1 (forearm) and subject #2 (forearm) and were programmed to deliver a consecutive train of negative (#1) and positive (#2) pulses. When the subjects are not touching, there is no performance change in the skinBLAST worn by the control subject. In contrast, when all three subjects hold hands or touch each other, the pulses are delivered effectively, changing the conductance of the skinBLAST worn by the control subject (see **Fig. 4g**). Additional sequences where the subjects do not touch one another can be found in Fig. S17.

In the future biomedical applications, such advanced neuromorphic tattoos can be used either locally (*e.g.,* wound healing) or globally (*e.g.,* in psoriasis and other skin diseases), aiding the development of self-aware stand-alone "smart" bioelectronic systems. Extending neuromorphic functionality from non-specific Nafion to ion-specific ionophores would bring extensive functionalities to respond to calcium and potassium ions that are vital to understanding human performance at the whole-body level. While modern state-of-the-art bioelectronic solutions always require an external system that records the sensed information, processes it, and sends back as a stimulation, we envision that skinBLASTs can bypass the cumbersome step. These neuromorphic tattoos can act as an intermediary self-aware sensors with a pre-formed threshold; when exceeding such threshold, the system can be triggered to release a counteragent (a chemical, heat, or electrical signal)[45] – for example for an immediate detection of epileptic seizures.

To summarize, we introduced three unique concepts. First – the concept of commercial Ag/AgCl gel electrodes as rapid and effective testbeds for 2D material characterization that do not require transfer. The solid electrolyte form of the gels allows them to apply a strong field-effect onto the wearable semiconductor. The gels can be used to characterize pure undoped and undamaged 2D materials, even on insulating substrates. Second – we reported on first-of-a-kind skin-wearable skin-gated transistor tattoos. Applying a field-effect tuning potential *via* the skin, we have effectively changed the conductivity level of graphene transistor tattoos, while the hypodermis play the role of electrolyte. Finally, we show that interbody neuromorphic communication is possible by employing the flexible neuromorphic tattoos that are controlled through the skin, opening on unique routes for advanced in-body neuromorphic computing systems.

## Methods

**Making GFET-Ts.** High-quality monolayer graphene was synthesized *via* the chemical vapor deposition (CVD) method on a copper foil substrate (Grolltex). The graphene/copper piece was then fixed onto a silicon (Si) wafer using Kapton tape. A layer of poly(methyl methacrylate) (PMMA, 950 A4) was spin-coated onto the sample at 2500 rpm for 60 seconds, followed by a 20-minute bake at 200°C on a hotplate. The copper substrate was removed by immersing the PMMA/graphene/copper sample in a 0.1 M $(NH_4)_2S_2O_8$ solution for 8-12 hours. The resulting PMMA-graphene layer was then cleaned in three stages of DI water rinsing (5 minutes, 5 minutes, and 30 minutes) to remove any residual copper etchant solution. The PMMA/graphene layer was subsequently transferred onto a target substrate. To make monolayer (1L) device, the target substrate is the temporary tattoo paper. A two-layer graphene sample was prepared by transferring the first PMMA-graphene piece on to the secondary graphene-copper sample; repeating the transfer steps to achieve 3L devices. Finally, the PMMA/graphene/tattoo paper was air-dried overnight at room temperature to ensure proper protection and support of the graphene layer during subsequent experimental procedures.

**Making skinBLAST.** The skinBLAST devices were fabricated by transferring graphene tattoos on top of ultrathin Nafion 211 membrane. Adhesive gold contacts are placed 2-10 mm apart from each other on top of the Nafion creating source and drain terminals. The graphene tattoos (~5-15 cm length and 3-10 mm width) are prepared for transfer onto the stack. The Nafion/graphene/PMMA stack is then supported by a skin-adhesive polymer (KRST tape or Tegaderm). When placed onto skin, it is Nafion that comes in direct contact with the skin, while graphene channel is sandwiched between the Nafion and PMMA/tape.

**Gels.** Five different gel electrodes, namely models 2330, 2360, 2560, and 2670-5 from 3M and Meditrace from Cardinal Health were used in this study. The gels were stored in ambient conditions. They were taped down to glass slides to ensure physical rigidity during use. Adhesive gold contacts are placed 2-10 mm apart from each creating source and drain terminals. Graphene tattoos were transferred on top of the devices with graphene facing down, hence coming in direct contact with the gels.

## 2D Materials.

**$MoS_2$**/isopropanol (IPA) ink solution was prepared through a molecular intercalation/exfoliation method [46]. The process utilized a two-electrode electrochemical cell comprising a thin cleaved $MoS_2$ crystal (molybdenite) as the cathode, a graphite rod as the anode, and tetraheptylammonium bromide (THAB, 98% from TCI) dissolved in acetonitrile (40 ml; 5 mg/ml or higher) as the electrolyte. In the intercalation step, a negative voltage (5-10 V) was applied to the $MoS_2$ cathode for 1 hour, facilitating the insertion of positively charged THA+ ions into the crystal and resulting in a fluffy bulk material. This material was then rinsed with absolute ethanol and subjected to sonication in a 40 ml 0.2 M PVP/DMF solution (PVP: molecular weight approximately 40,000, Sigma-Aldrich) for 30 minutes to achieve a greenish nanosheet dispersion. Subsequently, the dispersion underwent centrifugation and was washed with isopropanol (IPA) twice to eliminate excess PVP.

It is important to note that acetonitrile was used in our 1st step for the initial intercalation of THAB. Following the steps to fully disperse $MoS_2$ flakes, including 1 step of sonication in DMF and 3 steps of solvent exchange to remove unnecessary particles (organic and inorganic), the final product is MoS2 flakes dispersed in IPA, if there are any product from electrolysis of acetonitrile, it will be removed during these processes, assuring purity of the final material[22].

**PtSe$_2$** was prepared by the thermally assisted conversion (TAC) process. Initially, platinum (Pt) films with controlled thickness were deposited on a 2cm x 3cm $SiO_2$/Si wafer (300 nm $SiO_2$ thickness) or Kapton film, using the electron beam evaporator (Thermionics VE-100) at a fixed evaporation rate of 0.07 Å/s. Subsequently, the Pt-deposited sample was positioned in the central heating zone of a horizontal quartz tube furnace (Lindberg/Blue M Mini-Mite). In the upstream side of the furnace, we placed an alumina boat containing selenium (Se) precursors (99.9%, Millipore Sigma). The quartz tube was pumped down to a base pressure of ~30 mTorr and purged for 10 minutes with pure argon (Ar) gas to eliminate oxygen ($O_2$) and the other organic residue. The central area of furnace was then heated to 400°C over 50 minutes, maintaining this temperature for an additional 50 minutes under a continuous supply of Ar gas at a flow rate of 200 standard cubic centimeters per minute (sccm). The upstream area for the Se precursors was heated to ~200°C during the reaction. Following the reaction process, the quartz tube was naturally cooled down to room temperature.

**Graphene-on-Sapphire** samples were grown using atmospheric pressure chemical vapor deposition following a recipe outline elsewhere[24]. R plane sapphire wafers (from University Wafers) were diced into 5x10 mm strips and cleaned using acetone and IPA and dried with $N_2$. Following drying they were loaded in an annealing furnace for 4 hours at 1100°C to obtain a suitable surface reconstruction. The annealed sapphire was then loaded into a horizontal tube furnace. The temperature of the furnace was ramped to 1050°C; 200 sccm Ar, 80 sccm $H_2$ and 30 sccm $CH_4$ were flowed into the tube for 3 hours. Following growth, the gas flows were shut down and the furnace was cooled to room temperature.

**Electrical Characterization.** Electrical characterization was carried out using Keysight B2902A source-meter unit. Typically, the gate voltage was swept over a range of -0,9 to 0.9 V with $V_{DS}$ = 100 mV for gel and pig skin operation, and 40 mV for on-human operation. Compliance in drain-source and gate leakage currents was set to 50 µA for gel and pig skin experiments, and 1 µA for human experiments, to ensure that the operation potentials and currents are safe for humans. Conductance (S), defined as $I_{DS}/V_{DS}$, was used instead of direct current for proper benchmarking. Impedance and capacitance measurements were carried out using the LCR Meter (Hioki IM 3536). Each gel electrode was held at a constant voltage of 200 mV and swept from 4 Hz to 1 MHz with areas from 20, 40, 60, 80, and 100 mm$^2$.

**Porcine Skin.** Livestock animals derived porcine skin is used in the study, hence does not require IACUC approval. The hairless porcine skin is stored frozen (-20℃) and thawed for 2 hours at room temperature. Adhesive gold contacts are placed 2-10 mm apart from each other creating source and drain terminals. A gel electrode was placed on the skin to connect the gate terminal. The graphene tattoos (~5-15 cm length and 3-10 mm width) are prepared and transferred onto the skin. For long-term stability measurements, the skin underwent cycles of thawing-freezing.

**Human Subjects.** The human subject experiments were performed under the approval of the Institutional Review Board of the University of Texas at Austin (IRB no. 2018060058-MOD02). A total of N=12 presumably healthy subjects have been recruited to participate in this study.

**Characterizing skinBLAST.** The three-terminal devices were measured using Agilent B2902A. One of the SMUs is used to apply 0.1 V of the drain-source potential, while the gate is used in the current-pulsing mode to apply conductance changes. Two pulse train tests were used to characterize conductance change due to periodic pulsing. This was done by sending 50 consecutive write current pulses with corresponding width of 100 ms and amplitude of 1 μA. The pulse trains were conducted with both positive and negative pulses. The pulses were followed with 150-200 seconds relax to observe conductance retention. A pulse ramp experiment ran multiple sequences of a train of 20 negative pulses followed by a train of 20 positive pulses. The level experiment consisted of an arbitrary sequence of negative and positive current pulses of 10 μA that were pulsed through two gates every 1 second at a pulse width of 1 ms. Both tests were used to demonstrate distinct and repeatable conductance levels.

**Statistical tests** were performed using OriginPro. First, Levene's test for homogeneity of variances was applied to ensure that the assumption of equal variances across groups was met. A one-way analysis of variance (ANOVA) was subsequently used to compare the means of the groups. For pairwise comparisons, Tukey's Honestly Significant Difference (HSD) post-hoc test was conducted to identify significant differences between specific group means. Statistical significance was defined as $p < 0.05$ for all tests. OriginPro's built-in statistical tools were used to compute all p-values and post-hoc analyses.

**CrossSim simulator** was developed by Sandia National Laboratories as a device, circuit, and architecture simulator for crossbar arrays[47]. Shortly, the experimental ramping data (shown in **Fig. 4b**) is fed as an input into the simulator as a lookup table, where each update to each device in the network will be based on the experimental data. The ramping data not only includes the change in conductance per weight update, but also includes the cycle-to-cycle variability due to the inclusion of multiple ramping cycles. In this way, a relatively realistic estimate for the possible performance of the devices in a network can be drawn.

## Acknowledgments

DK, JAI, PV, and JB acknowledge funding from the US National Science Foundation grant CCF-FET under Grant No. 2246855. DA acknowledges the Cockrell Family Regents Chair Professorship. SL acknowledges funding from the US National Science Foundation graduate research fellowship under Grant No. 2021311125. JAI acknowledges the UT Austin Engineering Foundation Endowment. GCF acknowledges NSF award number EEC-1160494. Y.J. acknowledges the financial support from the US National Science Foundation (CAREER: 2142310).

## Author contributions

DK conceptualized the work and performed experiments with 2D-FETs. DK and NK performed the experiments with GFET-Ts. DK and GCF performed experiments with GFET-Ts stability. DK, SL, PV, and JB performed wearable neuromorphic experiments. SH and YJ provided $PtSe_2$ materials. DX and XD provided $MoS_2$ materials. SM provided graphene-on-sapphire samples. DA and JAC supervised the project. DK, NK, and SL wrote the original manuscript draft. All authors contributed to the manuscript review and editing.

## Competing interests

Authors declare that they have no competing interests.

## Data and materials availability

All data needed to evaluate the conclusions in the paper are present in the paper and/or the Supplementary Materials. Additional data related to this paper may be requested from the authors.

## Ethics statement

The human subject measurements were performed under the approval of the Institutional Review Board of the University of Texas at Austin (IRB no. 2018-06-0058).

## Additional information

Supplementary information is available for this paper online.

# Skin Controlled Electronic and Neuromorphic Tattoos

Dmitry Kireev[1,2*], Nandu Koripally[1,3], Samuel Liu[1], Gabriella Coloyan Fleming[4], Philip Varkey[1], Joseph Belle[1], Sivasakthya Mohan[5] Sand Sub Han[7], Dong Xu[8], Yeonwoong Jung[7,9,10], Xiangfeng Duan[11,12], Jean Anne C. Incorvia[1], Deji Akinwande[1,6,13]

[1] Chandra Department of Electrical and Computer Engineering, The University of Texas at Austin, Austin, TX, USA

[2] Department of Biomedical Engineering, University of Massachusetts Amherst, Amherst, Massachusetts, MA, USA

[3] Department of Electrical and Computer Engineering, University of California San Diego, La Jolla, CA, USA.

[4] Walker Department of Mechanical Engineering, The University of Texas at Austin, Austin, TX, USA

[5] Department of Materials Science and Engineering, The University of Texas at Austin, Austin, TX, USA

[6] Texas Materials Institute, University of Texas at Austin, Austin, TX, USA

[7] NanoScience Technology Center, University of Central Florida, Orlando, FL, USA

[8] Department of Materials Science and Engineering, University of California Los Angeles, Los Angeles, CA, USA

[9] Department of Materials Science and Engineering, University of Central Florida, Orlando, FL, USA

[10] Department of Materials Science and Engineering, University of Central Florida, Orlando, FL, USA

[11] Department of Chemistry and Biochemistry, University of California Los Angeles, Los Angeles, CA, USA

[12] California NanoSystems Institute, University of California Los Angeles, Los Angeles, CA, USA

[13] Department of Biomedical Engineering, The University of Texas at Austin, Austin, TX, USA

* Correspondence to: dkireev@umass.edu

**Table S1.** Comparing GFET-Ts to the state-of-the-art wearable neuromorphic skin-biased transistors

| Reference | Active Material | Operated through skin? | Max mobility, cm²V⁻¹s⁻¹ | Neuromorphic? |
|---|---|---|---|---|
| This work | Graphene<br>MoS₂<br>PtSe₂ | Yes | Graphene: ~6500<br>MoS₂: ~50<br>PtSe₂: ~1 | Yes |
| **Ref** [1] | DNTT | No | 0.41 | No |
| **Ref** [2] | PEDOT:PSS | No | <0.1 | No |
| **Ref** [3] | P3HT | No | 1.2 | No |
| **Ref** [4] | MoS₂ | Yes | --- | No |
| **Ref** [5] | p(g2T-T) | No | 1 | No |
| **Ref** [6] | N2200 | No | <0.2 | Yes |
| **Ref** [7] | p(g2T) | No | <0.3 | Yes |

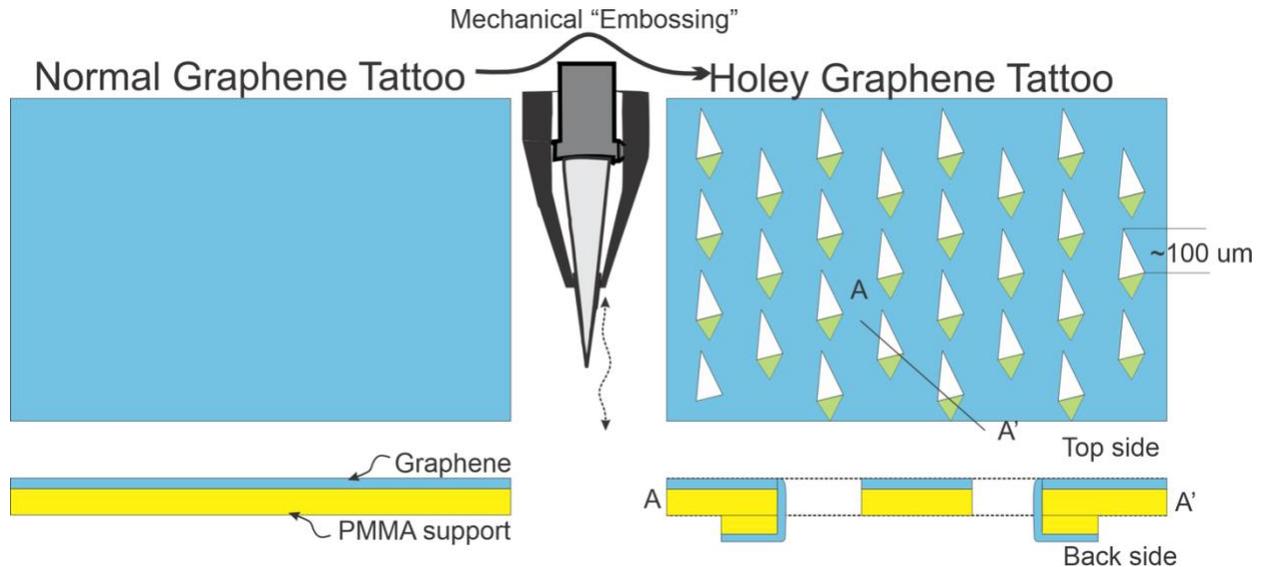

**Fig S1.** Holey graphene tattoos are created via mechanical embossing of the normal tattoos, creating triangular holes. Our experimental data suggests that while embossing, the punctured triangle-shaped pieces of the Graphene (blue)/PMMA(yellow) bend off, and when the tattoo is immersed into water (during transfer), the punctured pieces flap and attach to the back-side of the tattoo (supposed to be insulating). However, now there are regular elements of graphene sticking to the back side, establishing an effective electrical contact between the two sides of graphene tattoos. The holes are also essential for esurient water vapor transfer as has been shown in the previous work[8].

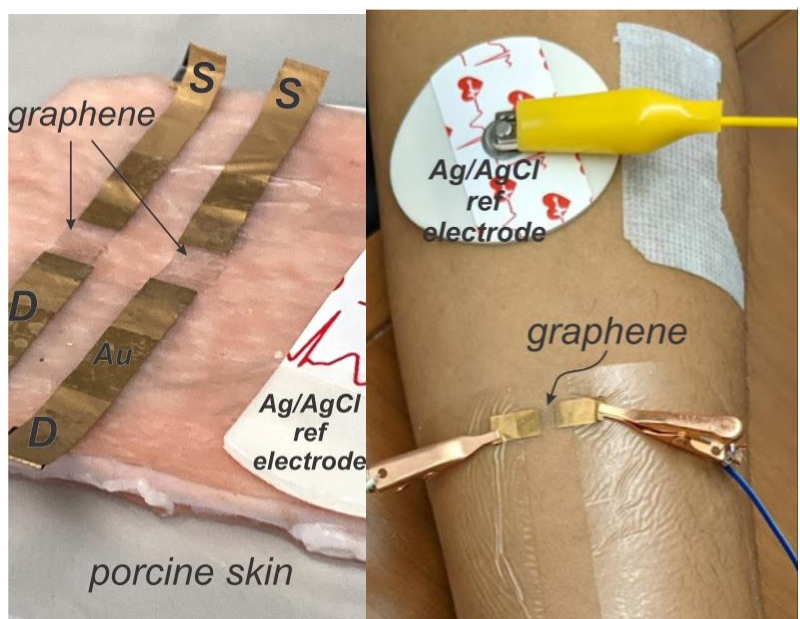

**Fig S2. Optical pictures of the GFET-Ts on porcine (left) and human (right) skin.** In both cases, the side Ag/AgCl is used as a reference electrode, graphene in the form of a GET placed on top of the skin, with graphene facing down and two conductive adhesive tapes are used as source and drain terminals.

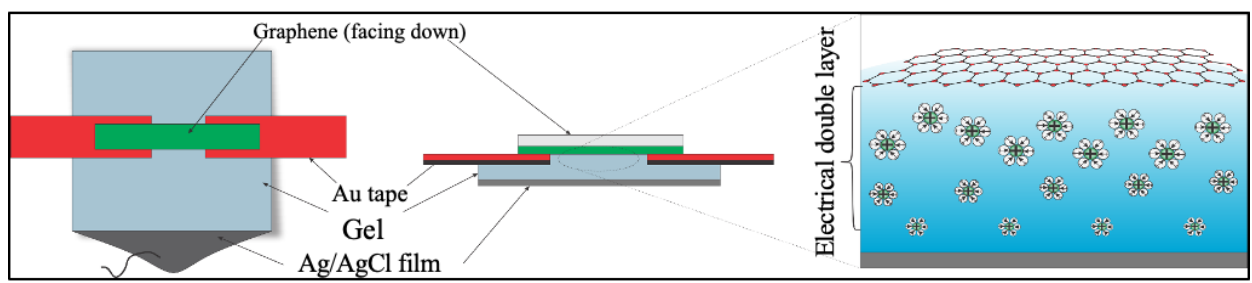

**Fig S3.** Schematic of the gel electrode as a test bed for 2D material characterization. The gel (blue gray) is supported by a backing layer with conductive Ag/AgCl that works as a reference electrode. On top of the gel, we place two stripes of Au/tape(adhesive) that works as source and drain leads. The distance between S and D defines the channel length, unless an additional patterning of the electrode is deposited onto the sample (see Fig. S10). Then, graphene or any other 2D material (green) is placed onto the gel, facing down, with sides connecting to S and D for electrical contacts.

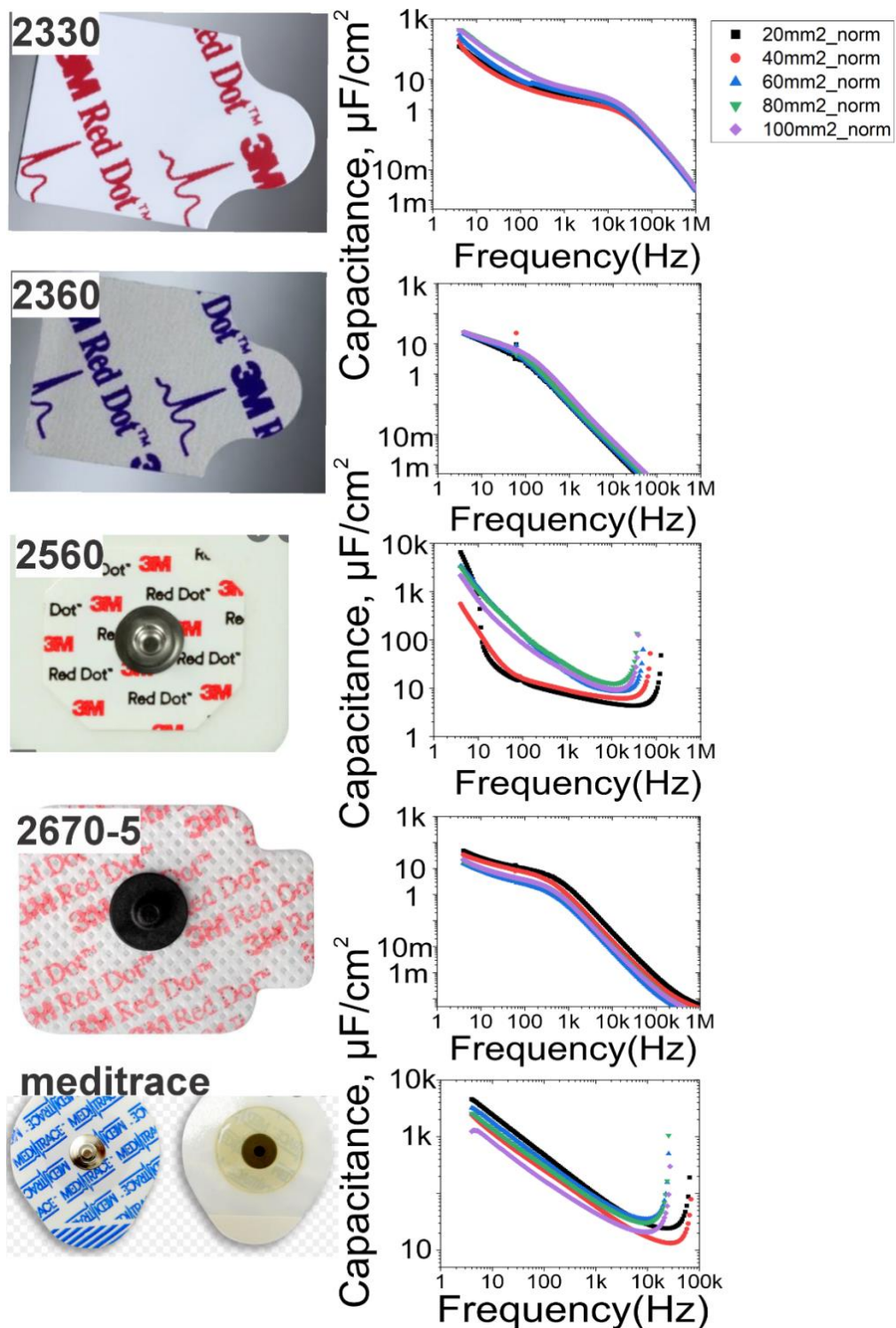

**Fig S4.** Characterization of the gels. Five different gel electrodes, models 2330, 2360, 2560, and 2670-5 from 3M and Meditrace from Cardinal Health were used in this study. Impedance and capacitance measurements on the gel electrodes were carried out using the LCR Meter (Hioki IM 3536). Each gel electrode was held at a constant voltage of 200 mV and swept from 4 Hz to 1 MHz with areas from 20, 40, 60, 80, and 100 mm$^2$.

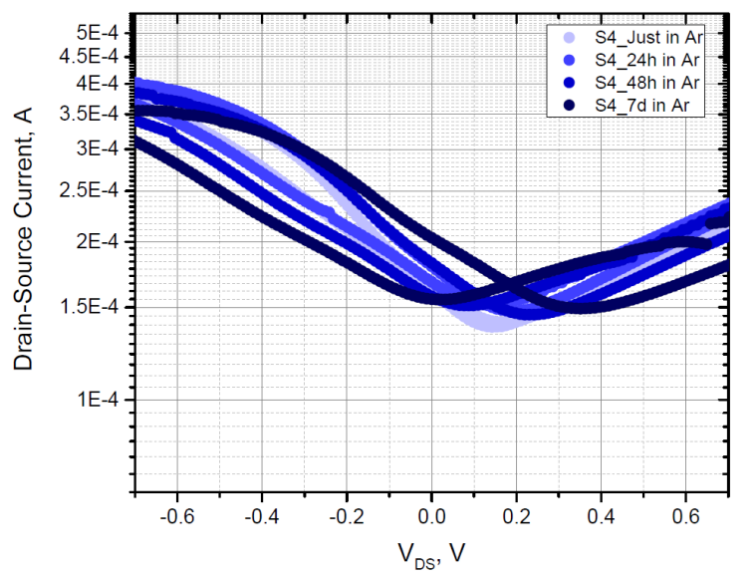

**Fig S5.** Ar Stability. A set of I-V curves of a GFET-T (#S4) that was stored in Ar atmosphere (glovebox) for 7+ days with no explicit degradation of the properties.

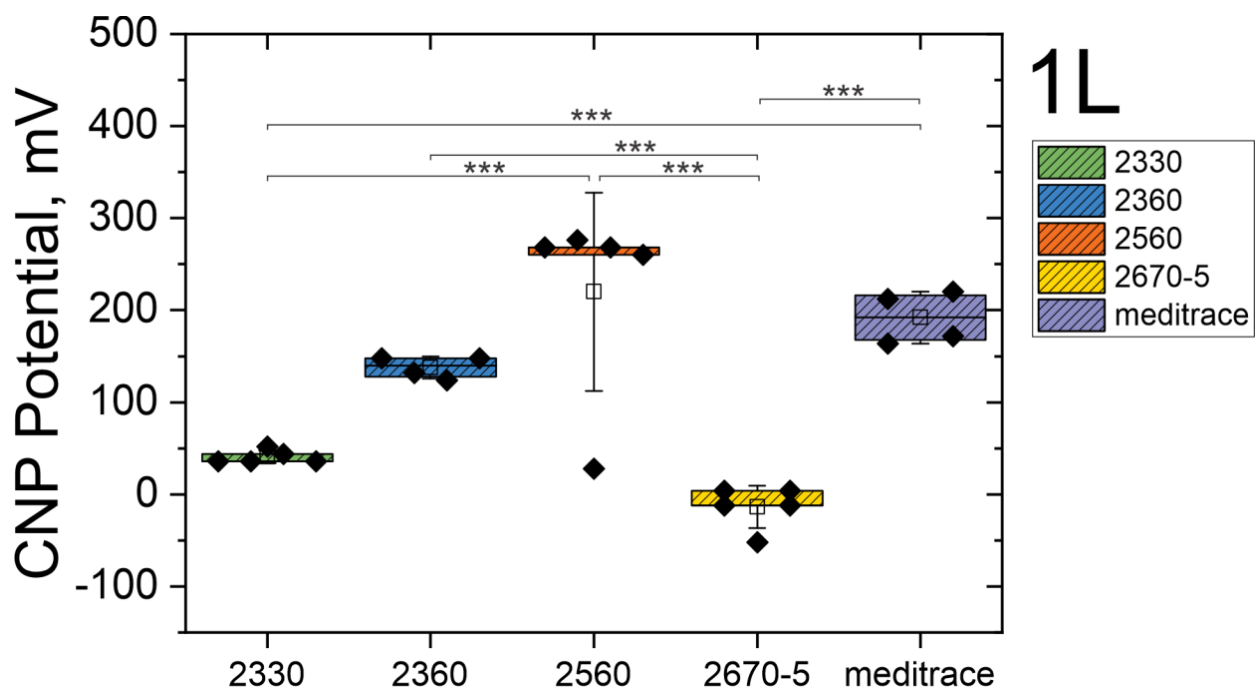

**Fig S6.** Statistical analysis of the graphene transistor performance (change in CNP potential) depending on the gel type. Statistical difference between two groups is denoted with * for (p<0.05), ** for (p<0.01) and *** for (p<0.005). Others are not significant (**n.s.**).

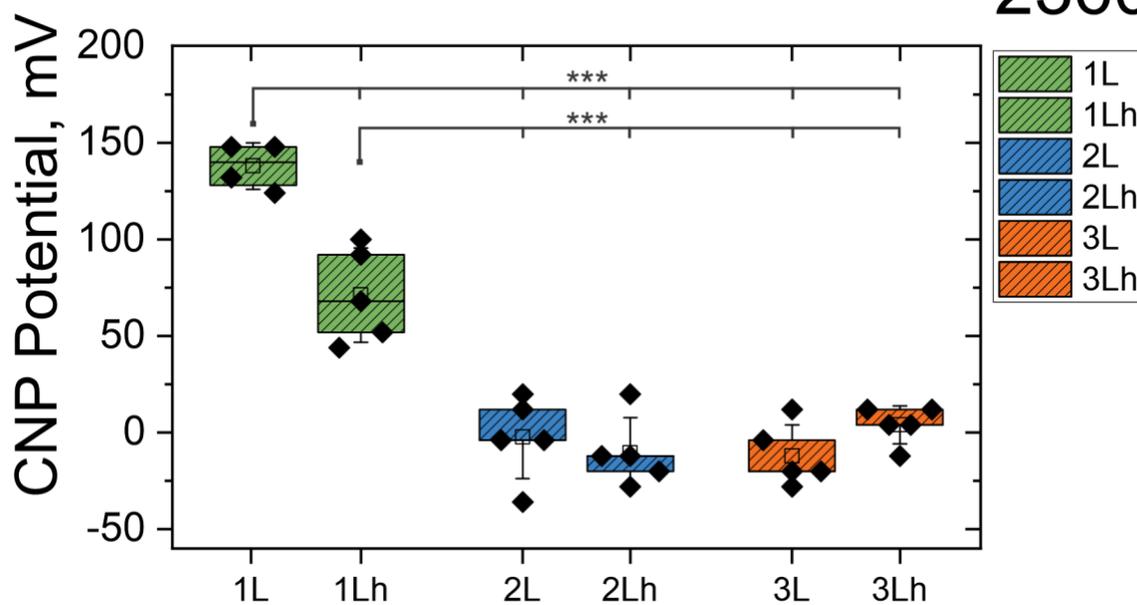

**Fig S7.** Statistical analysis of the graphene transistor performance (change in CNP potential) depending on the number of graphene layers. "h" here denotes the structures with micro-embossed holes (see Fig. S5). Hence, 1L and 1Lh only differ by the presence of an array of microholes in the tattoo structure. Statistical difference between two groups is denoted with * for (p<0.05), ** for (p<0.01) and *** for (p<0.005). Others are not significant (**n.s.**).

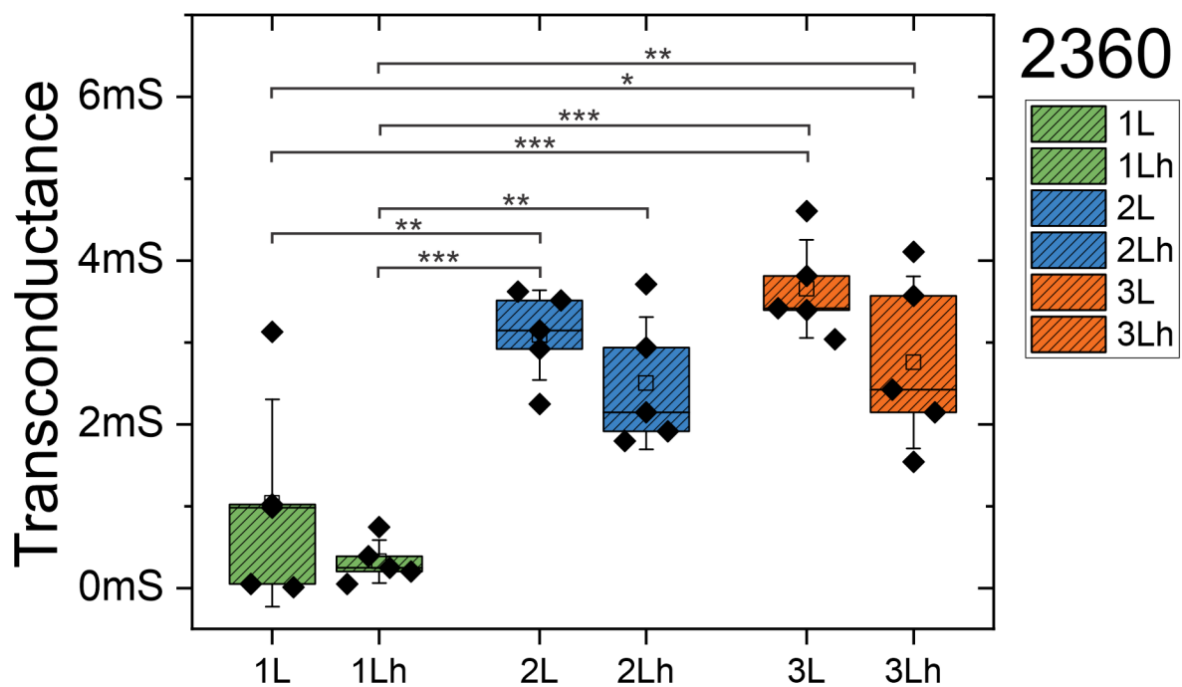

**Fig S8.** Statistical analysis of the graphene transistor performance (Transconductance) depending on the number of graphene layers. "h" here denotes the structures with micro-embossed holes (see Fig. S1). Hence, 1L and 1Lh only differ by the presence of an array of microholes in the tattoo structure. Statistical difference between two groups is denoted with * for (p<0.05), ** for (p<0.01) and *** for (p<0.005). Others are not significant (**n.s.**).

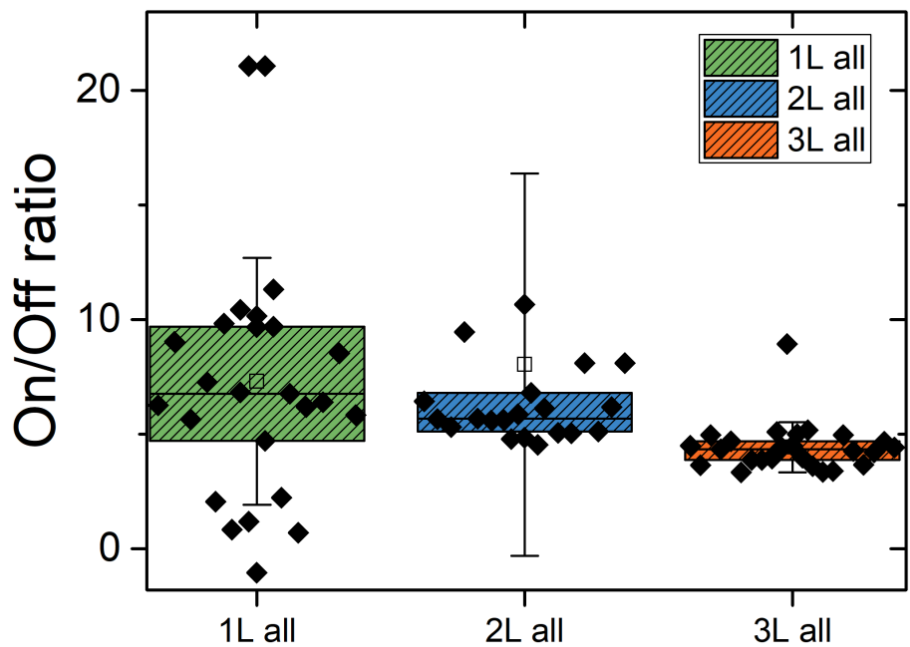

**Fig S9.** Statistical analysis of the graphene transistor performance (On/Off ratio) depending on the number of graphene layers. The groups are not significantly different (p>0.05).

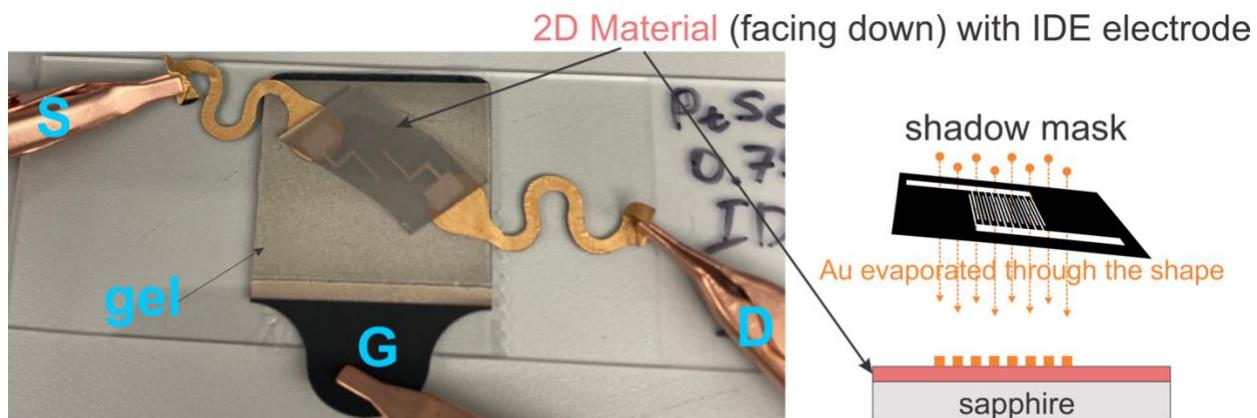

**Fig S10.** Optical photograph (left) and a schematic (right) of the specific design made to characterize $MoS_2$ spin-coated on flexible substrates or $PtSe_2$ samples grown on sapphire. Prior to the material placement on the gel electrodes, an IDE structure was deposited on top via a shadow mask.

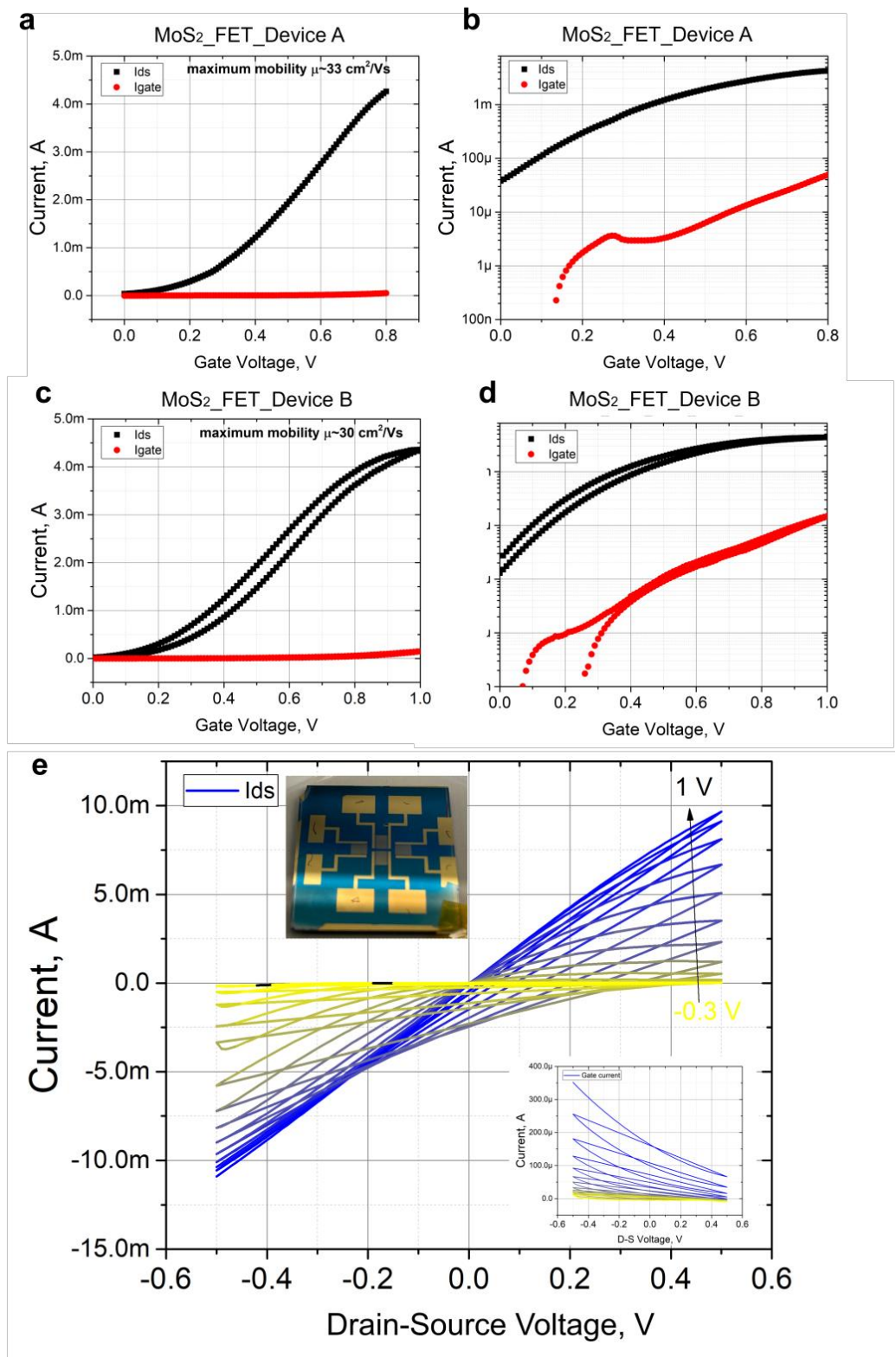

**Fig S11.** Additional transfer and output curves for two MoS₂ FET transistors measured *via* gel-gating.

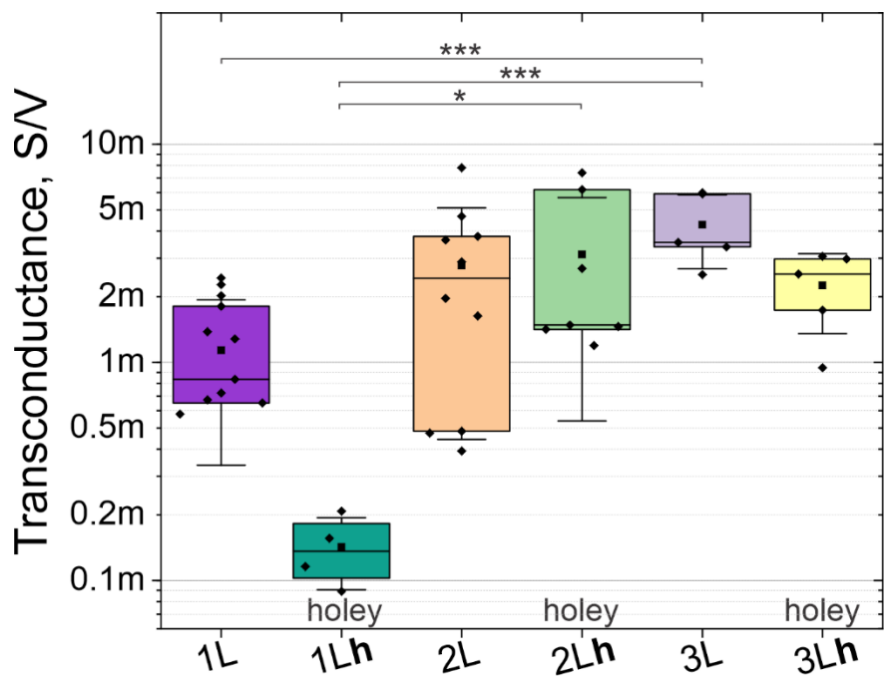

**Fig S12.** Transconductance of 6 different GFET-T types measured through porcine skin. "**h**" here denotes the structures with micro-embossed holes (see Fig. S1). Hence, 1L and 1Lh only differ by the presence of an array of microholes in the tattoo structure. Statistical difference between two groups is denoted with * for ($p<0.05$), ** for ($p<0.01$) and *** for ($p<0.005$). Others are not significant (**n.s.**).

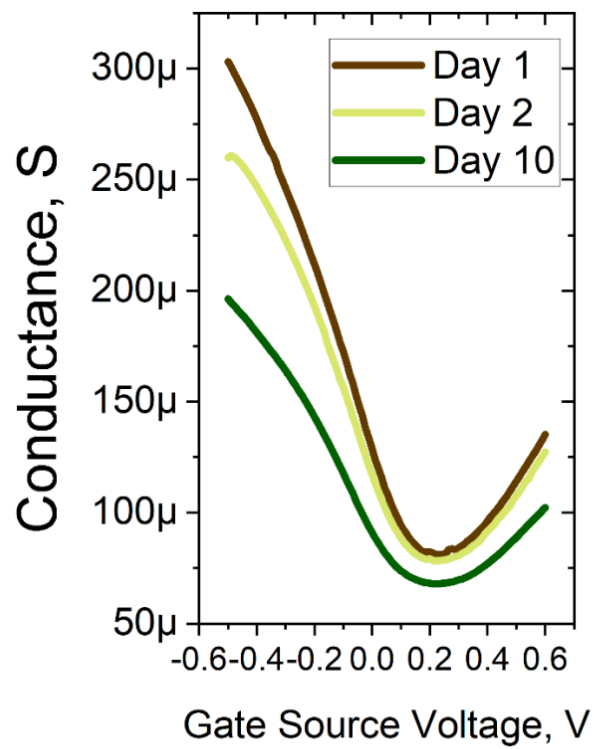

**Fig S13.** Performance (transfer curves) of a pig-skin GFET-T measured on days 1, 2, and 10. The sample is frozen in between the measurements and thawed at room temperature before re-measuring.

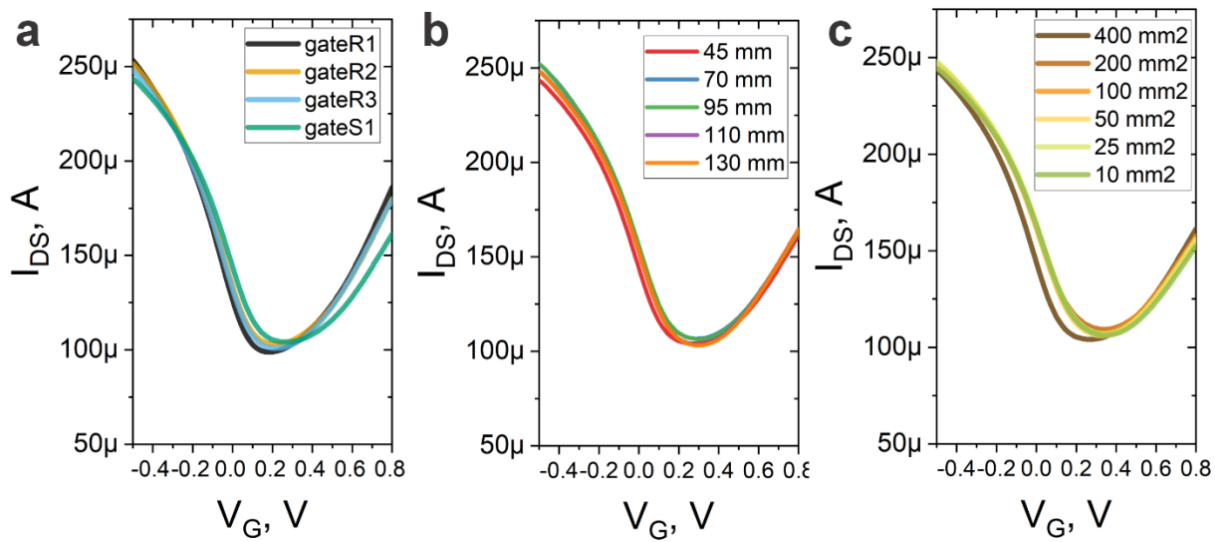

**Fig S14.** Performance of a GFET-T depending on the type of used reference electrode (a), the area of the reference electrode (b), and the distance to the reference electrode (c).

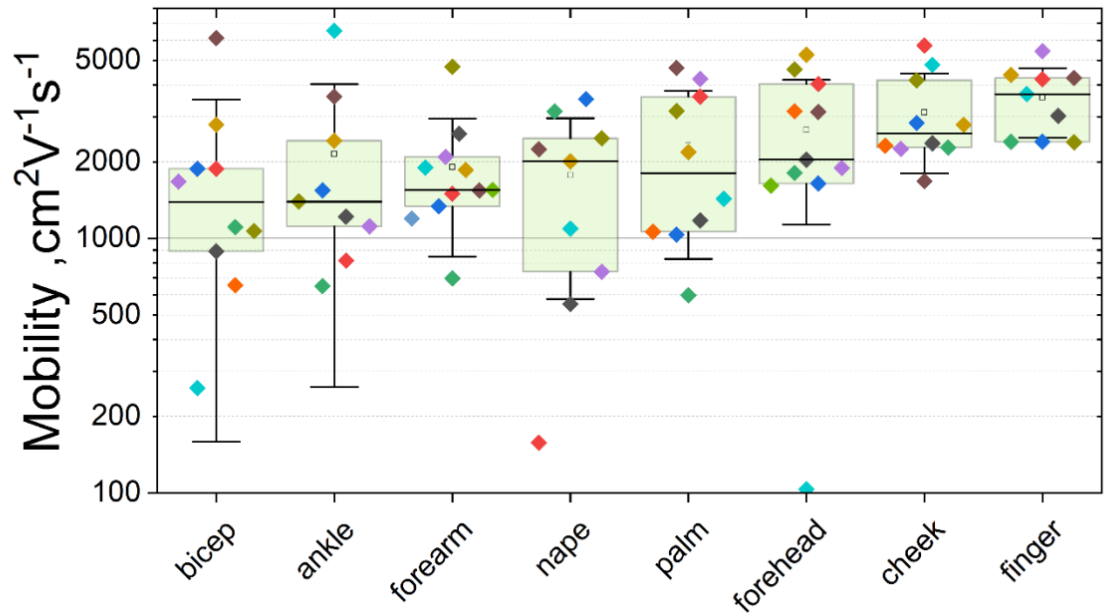

**Fig S15.** Statistical distribution of the mobility measured from 8 body locations and N=12 subjects. Each scatter color is associated with a subject, showing a significant trend for each subject. The box represents 25% and 75% with a mean; outliers are ±SD. The groups are not significantly different (p>0.05).

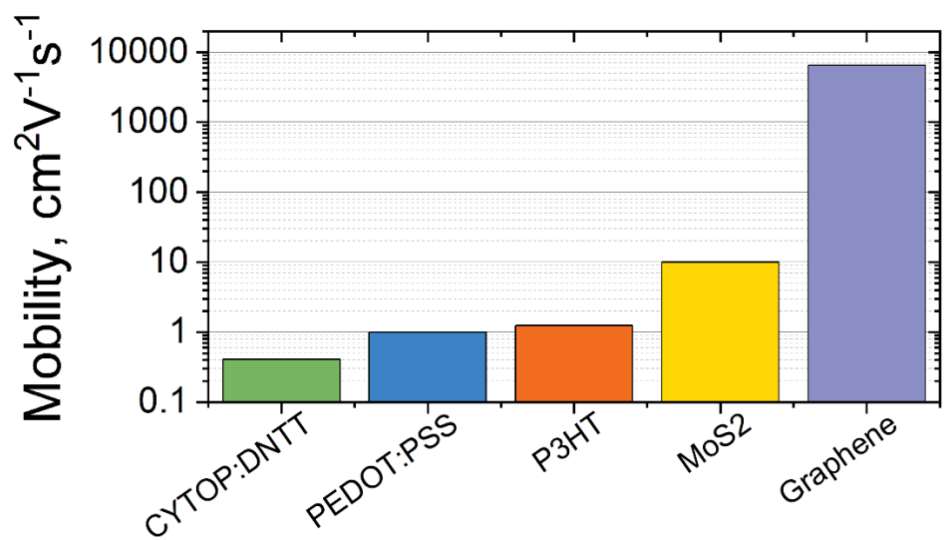

**Fig S16**. Benchmarking performance of the GFET-Ts in terms of field-effect mobility to other skin-interfaces wearable transistors: CYTOP:DNTT [1], PEDOT:PSS [2], P3HT [3], and MoS₂ [4]. Note – the y-axis is in log scale.

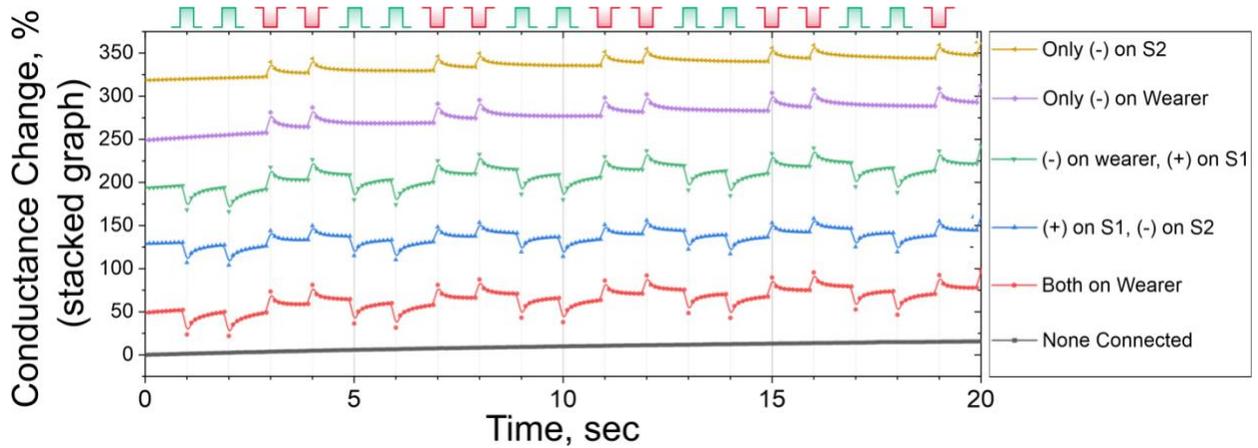

**Fig S17.** A series of sequences of neuromorphic performance of the devices. Note that the y-axis is stacked to accommodate the timetraces, hence the absolute values are arbitrary, but their relative amplitude is preserved. The neuromorphic device is always connected to the wearer. First, no gates were connected to the wearer – one can see (gray) that no conductance change. Then, in three consecutive measurements the gates were placed [A] both on the wearer (red), [B] (+) on subject 1, (-) on subject 2 (blue), and [C] (-) on the wearer, and (+) on subject 1 (green). All subjects held the hands on the wearer for these experiments. Clearly, the performance in such cases is identical. Then, it was only (-) connected to the wearer (purple) and (-) on subject 2, while subject 1 did not touch either of them (yellow).

**Table S2.** Human subjects measured and corresponding measurement locations.

| Subject | Bicep | Ankle | Forearm | Nape | Palm | Forehead | Cheek | Finger |
|---------|-------|-------|---------|------|------|----------|-------|--------|
| 1 | ✓ | ✓ | ✓ | ✓ | ✓ | ✓ | ✓ | ✓ |
| 2 | ✓ | ✓ | ✓ | ✓ | ✓ | ✓ | ✓ | ✓ |
| 3 | ✓ | ✓ | ✓ | ✓ | ✓ | ✓ | ✓ | ✓ |
| 4 | ✓ | ✓ | ✓ | ✓ | ✓ | ✓ | ✓ | ✓ |
| 5 | ✓ | ✓ | ✓ | ✓ | ✓ | ✓ | ✓ | ✓ |
| 6 | ✓ | ✓ | ✓ | ✓ | ✓ | ✓ | ✓ | ✓ |
| 7 | ✓ | ✓ | ✓ | ✓ | ✓ | ✓ | ✓ | ✓ |
| 8 | ✓ | ✓ | ✓ | ✓ | ✓ | ✓ | ✓ | ✓ |
| 9 | ✓ | ✓ | ✓ | ✓ | ✓ | ✓ | ✓ | ✓ |
| 10 | ✓ | -- | -- | -- | ✓ | ✓ | -- | -- |
| 11 | -- | -- | ✓ | -- | -- | ✓ | -- | -- |
| 12 | -- | -- | ✓ | -- | -- | -- | -- | -- |